\documentclass[prb,showpacs,floatfix,twocolumn]{revtex4}
\usepackage{epsfig}

\begin{document} 
\title{Fractional plateaus of the Coulomb blockade of coupled 
quantum dots}
 
\author{
Karyn Le Hur}


\affiliation{D\'epartement de Physique and CERPEMA,
Universit\'e de Sherbrooke, Qu\'ebec, Canada J1K 2R1.}

\newcommand{\br}{{\bf r}}
\newcommand{\ovl}{\overline}
\newcommand{\hw}{\hbar\omega}
\newcommand{\mybeginwide}{
    \end{multicols}\widetext
    \vspace*{-0.2truein}\noindent
    \hrulefill\hspace*{3.6truein}
}
\newcommand{\myendwide}{
    \hspace*{3.6truein}\noindent\hrulefill
    \begin{multicols}{2}\narrowtext\noindent
}

\begin{abstract}
Ground-state properties of a 
double-large-dot sample connected to a reservoir
via a single-mode point contact are investigated.
When the interdot transmission is {\it perfect} and the dots 
controlled by the same dimensionless gate voltage, we 
find that for any finite 
backscattering from the barrier between the lead and the left dot,
the average dot
charge exhibits a Coulomb-staircase behavior with 
steps of size e/2 and the capacitance peak period is
halved. The interdot electrostatic coupling here is weak. 
For strong tunneling between the left dot and 
the lead, we 
report a conspicuous intermediate phase in which the fractional plateaus 
get substantially altered by an increasing slope.
\end{abstract}

\maketitle

\section{Introduction}

At low temperature, the charge on an isolated metallic grain  
(micronmetric dot) is known to be quantized in 
units of the electron charge $e$. Even when the grain is weakly-coupled to a
bulk lead, so that electrons can occasionally 
hop from the lead to the dot and back, the 
grain charge remains to a large extent quantized\cite{Matv1}. This 
is commonly refered to as the Coulomb
blockade\cite{Grabert}. In the opposite limit of perfect 
transmission between the 
reservoir and the dot, the average dot charge now depends (in a
continuous manner) 
linearly on the applied gate voltage and the Coulomb blockade 
disappears\cite{no}. However, Matveev has shown that a 
crossover from the linear charge-voltage dependence to a 
{\it Coulomb-staircase} function occurs  
for any finite backscattering from the quantum point contact (QPC) 
between the grain and the lead\cite{Matveev}. The physics 
remains qualitatively
unchanged by increasing the reflection amplitude at the QPC.

Furthermore, close to the steps, 
the charge exhibits a nonanalytic logarithmic dependence on the voltage 
due to the presence of {\it two} spin channels entering the
dot, resulting in an underlying
two-channel Kondo model\cite{Nozieres}. 

Note also that the Coulomb blockade can be smeared out
by applying an in-plane magnetic field\cite{karyn2}. 

A direct measurement of the
average grain charge, has been made
possible using a single-electron transistor (SET) which has a sensitivity
well below a single charge as well as
a small input capacitance \cite{Grabert,Devoret}. 
In particular, some of the predictions above have been  
checked experimentally and its superiority to conductance measurements
of charge fluctuations demonstrated\cite{Berman}. 

{\it Here, 
we investigate exotic 
Coulomb staircases with fractional plateaus}. 

The simplest system we consider comprises two 
{\it large} symmetric dots, which can be viewed as an artificial molecule, 
connected to a single reservoir via a
single-mode QPC (Fig.~1). For a recent review on artificial molecules built
up with two dots, see Ref.~\onlinecite{Wilfred}.
Here, each dot is coupled 
with the same capacitance $C_{gd}$ to a side-gate. The term ``large dot''
implies that the spacing $\Delta\sim L^{-2}$ of the energy levels on each dot
vanishes compared to 
the dot's charging energy $E_c=e^2/(2C_{\Sigma})\sim L^{-1}$

\begin{figure}[ht]
\centerline{\epsfig{file=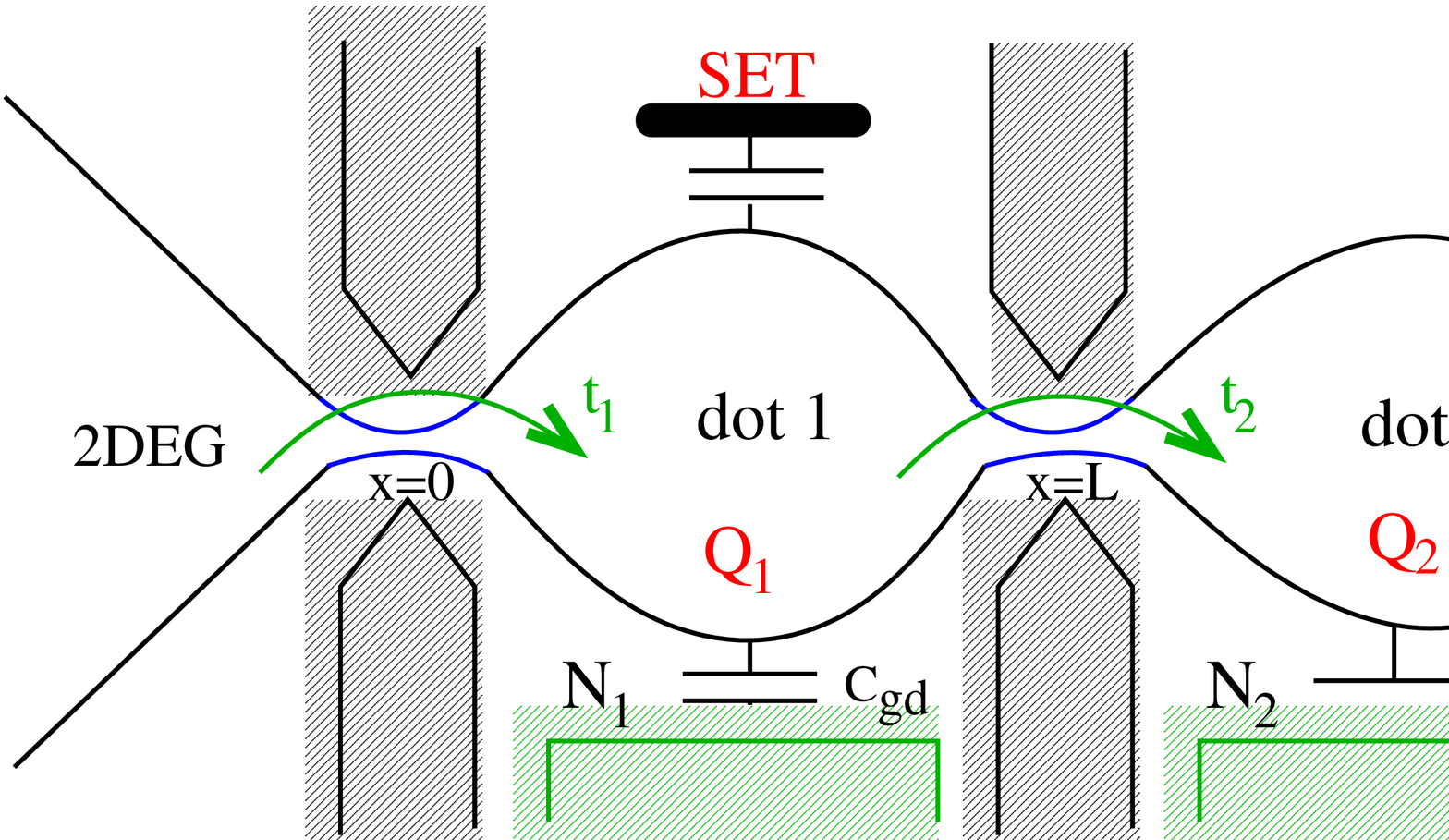,angle=0.0,height=4.2cm,width=
7.6cm}}
\vskip -0.05cm
\caption{A two-dimensional electron gas (2DEG) is coupled to two {\it large}
dots via a single-mode QPC. 
The number of electrons $Q_i$
on each dot is controlled via the dimensionless gate voltage $N_i$; The case
of interest here is $N_1=N_2=N$. 
The auxiliary gates
can be used to adjust the conductances at the QPCs. A SET may probe the 
average charge on a single dot.}
\end{figure}

\hskip -0.4cm where
$C_{\Sigma}\approx C_{gd}$. 
We already stress that
strong tunneling between the dots (``covalent binding'') 
is required in order to find
Coulomb staircases with fractional steps. 

For example, when the
interdot transmission is {\it perfect} and dot 1 is 
{\it weakly}-coupled to the lead, the interdot charge fluctuations are so
strong that only the total charge of the two dots, $eQ=e(Q_1+Q_2)$, 
can be {\it quantized} (but not the charge $eQ_i$ on an individual
dot). Thus, 
the electrostatic Hamiltonian of the two dots can be rewritten more
conveniently as (for details, see Refs.~\onlinecite{Matv2,Halperin}) 
\begin{equation}\label{Coulomb}\hskip -0.1cm
H_c[N]=\frac{E_c}{2}\hbox{\Large{(}}
Q-2N\hbox{\Large{)}}^2+2E_c\hbox{\Large{(}}Q_1-\frac{Q}{2}
\hbox{\Large{)}}^2-2E_cN^2. 
\end{equation}
The interdot capacitive coupling is {\it weak} in order to maximize the
interdot charge fluctuations\cite{note-time}.
Moreover, the symmetric dots are
controlled by the same gate voltage $V_G$ and 
$N=V_{G}C_{gd}/e$. From the electrostatic Hamiltonian, it can be easily 
inferred that the double dot behaves as
a single {\bf composite conductor} of quantized charge $eQ=2e\bar{Q}_1$
determined by the total gate voltage $2N$. When an 
electron tunnels into the left dot, i.e., $Q=1$, this implies that 
a charge $e$ is fluctuating 
back and forth between the dots and clearly
 $\bar{Q}_{i=1,2}[N]$ exhibits steps of size 1/2\cite{note-times}.
Moreover, close to a point $2N^*=(2n+1)/2$ $(n\in {\cal N})$, the charge
states with $Q=n\ (\bar{Q}_1=n/2)$ and $Q=n+1\ (\bar{Q}_1=n/2+1/2)$ are 
degenerate resulting in (sharp) peaks
in the single dot capacitance $C_1\propto\partial
\bar{Q}_1/\partial N$ (Fig.~2). 
Similar to the {\it conductance} peaks for two large dots tunnel-coupled to 
leads\cite{Matv2,Halperin}, we then observe that
strong interdot charge fluctuations produce the {\it halving} of the 
capacitance peak period. For an experimental proof, see e.g.
Ref.~\onlinecite{Waugh}

Based on two-impurity two-channel Kondo models (2CKMs) (small 
dots coupled to leads are described by a two-impurity 1CKM\cite{georges}), 
below we thoroughly analyze the evolution of
the fractional steps as a function of the hopping parameters
$t_1$ and $t_2$ (Fig.~1). Some aspects of the problem will join up with 
previous works on the conductance through a double (large) dot 
structure\cite{Matv2,Halperin}.

From here on, we assume that a single orbital channel with two 
spin polarizations $\alpha=\ \uparrow,\downarrow$ enters the 
double dot. Again, we assume that the level spacing on each dot (almost)
vanishes which means that we consider
a {\it continuous} spectrum in each dot and we neglect the
mesoscopic corrections to the capacitance $C_{gd}$; the size of a dot 
can thus exceed the effective Bohr radius 
($\sim \mu m$ in Refs. \onlinecite{Berman,Waugh}). Temperature 
will be taken to be zero $(T=0)$.

\section{Weak coupling with lead}

Weak tunneling $(t_1\ll 1)$ between the
lead and the composite dot produces corrections to the Coulomb 
staircase behavior found above. 

More precisely, for perfect interdot transmission $(t_2\rightarrow 1)$, we 
can describe the {\it composite dot} in the vicinity of the two QPCs
by the same field operator $\Psi_{c\alpha}(x)$.
Additionally, close to a degeneracy point 
$N^*=(2n+1)/4$, only the states with $Q=n$ and
$Q=n+1$ are allowed and, thus, following Ref.~\onlinecite{Matv1}, the tunneling
Hamiltonian for this truncated system takes the form:
\begin{equation}\label{tunnel}
H_{t}=\sum_{\alpha}\hbox{\huge{(}}
t_1\Psi^{\dagger}_{c\alpha}(0)\Psi_{r\alpha}(0)S^{+}+h.c.\hbox{\huge{)}}.
\end{equation}
$\Psi_{r\alpha}$ stands for the electron operator in the {\it lead},
and the 
spin operator $S^+$ guarantees that when an electron tunnels into the 
{\it double dot}, the total
charge $Q$ only changes from $n$ to $n+1$\cite{Matv1,karyn2}; we then have the 
 equalities\cite{note2} 
\begin{equation}\label{merge}
\bar{Q}= 2\bar{Q}_1 
=(n+1/2)+\bar{S}_{z}.
\end{equation} 
Following the route of the
single-dot problem\cite{Matv1,karyn2}, now we can identify $
s_{\alpha}^{-}(0)=\Psi^{\dagger}_{c\alpha}(0)\Psi_{r\alpha}(0)$ as an
electron pseudo-spin operator acting on the (orbital) indices
$j=r,c$ and finally recover a 2CKM\cite{Nozieres,Tsvelik}. The two channels 
are 
the two spin states of an electron. In particular, 
Eq.~(\ref{Coulomb}) can be viewed as a local
magnetic field $hS_{z}$ with $h\propto (2n+1-4N)$. This results in
\begin{equation}\label{merge2}
\bar{Q}_1 - \frac{2n+1}{4} \propto (2n+1-4N)\ln 
\hbox{\large{(}}|N-\frac{2n+1}{4}|
\hbox{\large{)}}.
\end{equation}
To sum it up, we recover a standard logarithmic 
form 
\begin{equation}
\delta C_1=C_1-C_{gd}
\propto -\ln\hbox{\large{(}}|N-\frac{2n+1}{4}|\hbox{\large{)}},
\end{equation}

\begin{figure}[ht]
\centerline{\epsfig{file=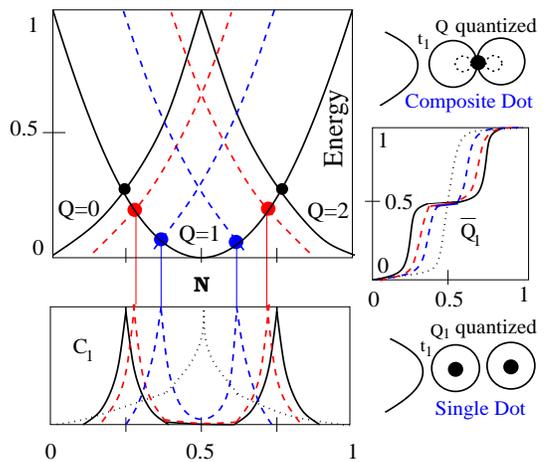,angle=0.0,height=6cm,width=7cm}}
\caption{Charging energies $(+\delta E)$ 
of the ``composite'' dot as a function of $N$ given 
in units of $E_c$; $t_1$ is {\it small}. Each eigenstate with $Q=n$
  gives 
rise to a parabola. 
The solid lines correspond to $r_2=0$ and dashed lines to increasing $r_2$
couplings. For $r_1=1-t_1\rightarrow 1$, $Q$ is
  quantized and for symmetric dots 
this guarantees $\bar{Q}_1=n/2$
until $r_2\rightarrow 1$ (Eqs.~(\ref{steps}),(\ref{splits2})).}
\end{figure}

\hskip -0.4cm for the capacitance peaks. 

We now discuss the situation in which the interdot tunneling is
strongly decreased, $(t_1;t_2)\ll 1$. Each dot 
is described by its own operator $\Psi_{i\alpha}$ and the
Coulomb term should be written in a more common way 
as\cite{note-time,Halperin}:
\begin{equation}\label{Coulomb2}
H_c^1[N]+H_c^2[N]=
E_c\sum_{i=1,2}\hbox{\Large{(}}Q_i-N\hbox{\Large{)}}^2-2E_c N^2.
\end{equation}
When $t_2\rightarrow 0$, we converge to a {\bf single-dot} 
problem\cite{Matv1,karyn2}: $Q_1$ is quantized and
we could not use $\bar{Q}_1\sim Q/2$ in Eq.(\ref{Coulomb})!
The degeneracy points now occur for 
$N^*_s=(2k+1)/2$ $(k\in {\cal N})$ 
and obviously the period of the capacitance peaks then {\it doubles} (Fig.~2). 

As soon as $t_2$ is finite $(t_2\sim t_1)$ and $N\approx 1/2$, we propose to
modify the tunneling Hamiltonian as:  
\begin{eqnarray}\hskip -1.5cm
H_{t}\hskip -0.1cm &=&\hskip -0.1cm\sum_{\alpha} \hbox{\huge{(}}
t_1\Psi^{\dagger}_{1\alpha}(0)\Psi_{r\alpha}(0)+t_2
\Psi^{\dagger}_{2\alpha}(L)\Psi_{1\alpha}(L)+h.c.\hbox{\huge{)}}
\hskip 0.3cm\\ \nonumber
\hskip -0.1cm&=&\hskip -0.1cm\sum_{\alpha} 
\hbox{\huge{(}}
 t_1 s_{\alpha}^-(0)S_1^{+}+
 t_2 s_{\alpha}^-(L)S_2^{+}+h.c.
\hbox{\huge{)}},
\end{eqnarray}
where $S_1^{+}(S_2^{+})$ emphasizes that the charge on dot 1(2) 
only changes from $0$ to $1$; For more details, see note~\onlinecite{note33}. 

Again, the index $j=1,2,r$
-which designates the location of an electron in the setup- in the 
$\Psi_{j\alpha}$ operator can mimic an internal ``orbital'' degree of freedom.
It is then straightforward to define two spin operators
at $x=(0,L)$ acting on the orbital space, similar as in 
Ref.~\onlinecite{Matv1}:
\begin{eqnarray}
s_{\alpha}^-(0) &=& \Psi^{\dagger}_{1\alpha}(0)\Psi_{r\alpha}(0) \\ \nonumber
s_{\alpha}^-(L) &=& \Psi^{\dagger}_{2\alpha}(L)\Psi_{1\alpha}(L).
\end{eqnarray} 
This two-impurity (two-channel) Kondo model is particularly 
convenient to revisit the behavior
of charge fluctuations close to the degeneracy points $N_s^*$; the crucial 
point being that a finite bare coupling
$t_2$ (like $t_1$) will be strongly renormalized at low 
temperatures\cite{georges2}.

At the fixed point $(T=0)$ and, e.g., close to the {\it degeneracy} point 
$N^*_s=1/2$, the two dots 
will {\it merge} into one and therefore by analogy to Eq.~(\ref{merge}) 
we must correctly reidentify 
\begin{equation}
\bar{Q}=(j+1/2)+\bar{S}_{z}=2\bar{Q}_1,
\end{equation}
where $j=(0;1)$\cite{note3}. Moreover, the Coulomb term in the 
fixed-point basis takes the form
$hS_{z}$ where $h\propto (1-2N\mp 2\kappa{\cal T}_2)$ for $j=(0;1)$; 
${\cal T}_2=(t_2)^2$ and $\kappa>0$. Away from the point 
$N^*_s=1/2$, second order perturbation theory in $t_2$ is
accurate, and we have taken into account the relative energy shift 
between {\it even} and {\it odd} Q-states
\cite{Matv2,Halperin}: 
\begin{equation}
\delta E\propto -4{\cal T}_2\ln 2.
\end{equation}
Similar to Eq.~(\ref{merge2}),  we are thus led to (for $j=0,1$ respectively)
\begin{equation}\label{steps}\hskip -0.3cm
\bar{Q}_1 =\left\{
\begin{array}{r@{\quad}l}\hskip -0.1cm
\frac{1}{4}-b\hbox{\Large{(}}N-\frac{1}{2}+\kappa{\cal T}_2
\hbox{\Large{)}}\ln
\hbox{\Large{(}}|N-\frac{1}{2}+\kappa{\cal T}_2|\hbox{\Large{)}},\\
\frac{3}{4}-b\hbox{\Large{(}}N-\frac{1}{2}-\kappa{\cal T}_2
\hbox{\Large{)}}\ln
\hbox{\Large{(}}|N-\frac{1}{2}-\kappa{\cal T}_2|\hbox{\Large{)}}.
\end{array} \right. \hskip -0.2cm
\end{equation}
$b>0$ is a parameter which is inversely proportional to the Kondo energy scale.
By continuity, a tiny step appears at $\bar{Q}_1=1/2$, and the
single-dot capacitance peaks are already split by $\sim 2\kappa
{\cal T}_2$ (Fig.~2). 

The progressive pairing of the capacitance sub-peaks
close to $t_2=1$ will be studied later (Eq.~(24)).

\section{Strong coupling with lead}

Now, we mainly consider the case
where all the junctions have conductances close to $2e^2/h$,
i.e., reflection amplitudes are small $(r_1;r_2)\ll 1$. 

In this case, the whole system can be viewed as a single conductor and,
for convenience, we will use the unique field operator 
$\Psi_{r\alpha}(x)$\cite{extra}. We can write 
$\Psi_{r\alpha}(x)= {\rm exp}(ik_Fx)\Psi_{+\alpha}(x)+{\rm exp}
(-ik_Fx)\Psi_{-\alpha}(x)$,
$\Psi_{+\alpha}$ and $\Psi_{-\alpha}$ 
describe right- and left-moving fermions respectively. The kinetic 
energy obeys
\begin{equation}
H_{k}=i v_F\int_{-\infty}^{+2L}dx\ \hbox{\Large{(}}
\Psi_{+\alpha}^{\dagger}\partial_x\Psi_{+\alpha}-
\Psi_{-\alpha}^{\dagger}\partial_x\Psi_{-\alpha}\hbox{\Large{)}}, 
\end{equation}
$v_F$ being the Fermi velocity.
The backscattering term(s) takes the standard form:
\begin{equation}\label{bs1}
H_b=v_F\sum_{\alpha}\hbox{\Large{(}}r_{1,2}\Psi^{\dagger}_{+\alpha}(0,L)
\Psi_{-\alpha}(0,L)+h.c.\hbox{\Large{)}},
\end{equation}
and interactions in a grain are 
embodied via the general 
Coulomb Hamiltonians $H_c[N_1]+H_c[N_2]$, in Eq.~(\ref{Coulomb2}).

At low energy, we proceed with this model by bosonization of the
one-dimensional Fermi fields\cite{karyn2}.
In those variables, the kinetic energy
yields a separation of the spin and charge and the resulting Hamiltonians have
plasmon-like excitations. Here, $\partial_x\phi_j$ with $j=(c,s)$ measures 
fluctuations of charge/spin density
and $\Pi_j=\partial_x\theta_j$ being its conjugate momentum. The Coulomb
Hamiltonians take the forms (We could also employ Eq.~(\ref{Coulomb})):
\begin{eqnarray}
H_c^1[N_1] &=&
\frac{2E_c}{\pi}\hbox{\huge{(}} \phi_c(0)-\phi_c(L)-\sqrt{\frac{\pi}{2}}
N_1\hbox{\huge{)}}^2-E_c{N_1}^2 \hskip 0.7cm\\ \nonumber
H_c^2[N_2] &=& \frac{2E_c}{\pi}\hbox{\huge{(}} \phi_c(L)-\phi_c(2L)
-\sqrt{\frac{\pi}{2}}N_2\hbox{\huge{)}}^2-E_c{N_2}^2.
\end{eqnarray}
To minimize the Coulomb energies
when the transmissions at the two QPCs are {\it both} perfect, 
we easily recover that the dot's charges evolve
continuously (linearly) as a function of the gate 
voltages\cite{no,Matveev}:
\begin{eqnarray}
\bar{Q}_{1} &=& \frac{\sqrt{2}}{\pi}\hbox{\Large{(}}\phi_c(0)-\phi_c(L)
\hbox{\Large{)}}=N_1 \\ \nonumber 
\bar{Q}_{2} &=& \frac{\sqrt{2}}{\pi}\hbox{\Large{(}}
\phi_c(L)-\phi_c(2L)\hbox{\Large{)}}=N_2.
\end{eqnarray}
Remember that for $r_1=r_2=0$, the Coulomb blockade physics is totally
suppressed. In our geometry there are no charge fluctuations 
at $x=2L$ and then $\phi_c(2L)=cst$.

Furthermore, following the traditional 
route of the single-dot problem for this regime\cite{Matveev,karyn2}, 
the backscattering term may be rewritten as
\begin{eqnarray}
\label{Emery}
\hskip -0.2cm
H_{b}\hskip -0.1cm &=&\hskip -0.1cm\frac{\sqrt{\gamma a E_c v_F}}{\pi a}4r_1
\cos(\pi(N_1+N_2))
\cos\left(\sqrt{2\pi}\phi_s(0)\right){\cal T}_{1x}
\hskip 0.7cm\\ \nonumber
&+&\hskip -0.1cm
\frac{\sqrt{\gamma a E_c v_F}}{\pi a}4r_2\cos\left(\pi N_2\right)
\cos\left(\sqrt{2\pi}\phi_s(L)\right){\cal T}_{2x}. \nonumber
\end{eqnarray}
Since the charge 
fluctuations on each dot cannot depend 
 on the precise size of a dot, we
must equate $\phi_c(2L)=2k_F L/\sqrt{2\pi}$ and rescale
$\phi_c(0)\rightarrow \phi_c(0)+2k_F L/\sqrt{2\pi}$.
Here $\gamma$ obeys $\gamma={\rm e}^{\cal C}$ where ${\cal C}
\approx 0.5772...$ is the 
Euler-Mascheroni constant and $a$ is a short-distance cutoff. 
We have introduced two {\it commuting} impurity spins ${\cal T}_{1}$
and ${\cal T}_{2}$ (which here are {\bf not} related to the charge on 
each dot). Clearly, the 
${\cal T}_{1x}$ and ${\cal T}_{2x}$ spin-operators 
both commute with the Hamiltonian and must be simply 
identified as c-numbers, i.e., 
${\cal T}_{1x}=1/2$ 
(or $-1/2$) and similarly for ${\cal T}_{2x}$. 
Eq. (\ref{Emery}) must be viewed as an extension of the 2CKM at the 
Emery-Kivelson lign\cite{Emery2}. 

To compute the correction to the average 
dot charge(s), here we must ``de-bosonize'' the problem as\cite{Matveev,karyn2}
\begin{eqnarray}
H_{b} &\approx& \frac{iJ_{1x}}{\sqrt{4\pi a}}\hbox{\Large{(}}\psi(0)+
\psi^\dagger(0)\hbox{\Large{)}}{\zeta}_1
\\ \nonumber
&+& \frac{iJ_{2x}}{\sqrt{4\pi a}}\hbox{\Large{(}}
\psi(L)+\psi^\dagger(L)\hbox{\Large{)}}{\zeta}_2.
\end{eqnarray}
${\zeta}_1$ and ${\zeta}_2$ are two Majorana fermions, and the Kondo
exchanges above read $J_{1x}=4r_1\sqrt
{a\gamma E_c v_F}\cos(\pi(N_1+N_2))$ and $J_{2x}=4r_2\sqrt
{a\gamma E_c v_F}\cos(\pi N_2)$. In
the absence of an applied magnetic field, there is no net magnetization and no
spin current on the whole region $[-L;L]$ and, thus, we have 
approximated\cite{extra}
$\psi(L)\approx \exp(i\sqrt{2\pi}\phi_{s}(L))$ (For more

\begin{figure}[ht]
\centerline{\epsfig{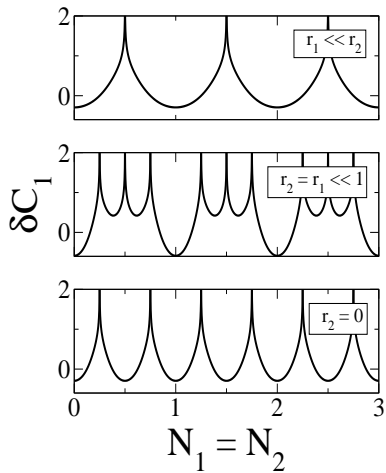}}
\caption{Dot's 
differential capacitance for {\it small} $r_1=0.4$.
For $r_2\approx 0$ the system 
behaves as a single {\it composite conductor} whereas
for $r_2\rightarrow 1$ 
we must recover a {\it single-dot} problem. For
$r_2\approx r_1\ll 1$ the system cannot decide between
  those two ground
states giving a ``3-peak'' capacitance profile, i.e., unstable fractional
steps (Fig.~4 and Eq.~(\ref{comp})): Charge fluctuations are
important at $N=1/4$ and at $N=1/2$ as well.}
\end{figure}

\hskip -0.4cm 
explanation, see Ref.~\onlinecite{karyn2}). 
The fermionic model here 
generates two Kondo resonances
\begin{eqnarray}\label{energies}
\Gamma_1&=&\frac{{J_{1x}}^2}{4 \pi a v_F}
=\frac{E_c\gamma}{\pi}(2r_1)^2\cos^2(\pi(N_1+N_2))\\ \nonumber
\Gamma_2&=&\frac{{J_{2x}}^2}{4 \pi a v_F}
=\frac{E_c\gamma}{\pi}(2r_2)^2\cos^2(\pi N_2),
\end{eqnarray}
and all the quantities of interest will be now inferred from
the quantum correction of the ground state energy
\begin{equation}\label{energy}
\delta E= -\frac{\Gamma_1}{\pi}\ln(E_c/\Gamma_1) -
\frac{\Gamma_2}{\pi}\ln(E_c/\Gamma_2), 
\end{equation}
which implies that impurities are 
{\it independently} screened.
Let us discuss the case of symmetric dots: $N_1=N_2=N$. 
The correction
to the average charge on each dot $\delta\bar{Q}_i$  and the dot's differential
capacitance $\delta C_i$ obey: $\delta\bar{Q}_i=\bar{Q}_i-N\propto
-\partial\delta E/(E_c 
\partial N)$ and $\delta C_i\propto \partial \delta\bar{Q}_i/\partial N$.
{\it For clarity's sake, results have been summarized in Figs.~3,4}.

\begin{figure}[ht]
\centerline{\epsfig{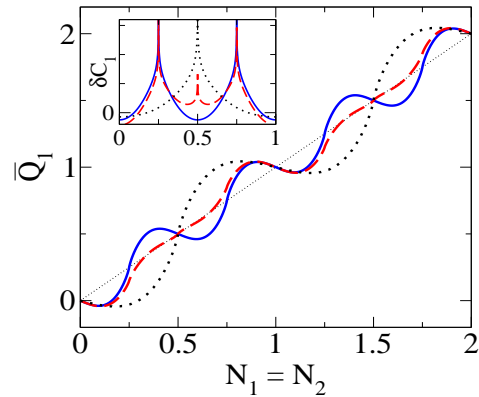}}
\vskip -0.2cm
\caption{Evolution of the fractional plateaus for {\it small} $r_1$. (Note 
e.g. that for $r_2=0$ we have only taken into account 
the main leading term when taking the derivate of Eq. (19), which explains
the ``slightly'' negative slope in the middle of a plateau).
The solid line is for $r_1=0.2$
and $r_2=0$, the dashed line for $r_1=r_2=0.15$ (fractional plateaus
now acquire a {\it positive} slope) and the dotted line
for $r_1=0.3$ and $r_2\rightarrow 1$.}
\end{figure}

\subsection{$r_2\rightarrow 0$}

For a double dot connected by a reflectionless constriction
$0\leftarrow r_2\ll r_1(\ll 1)$, using the formulas above, we easily
recover fractional charge plateaus
with steps 1/2 and capacitance peaks with halved period. Again, the
double dot behaves as a single {\bf composite conductor} of {\it
 quantized} charge $Q\approx 2\bar{Q}_1$. In particular, we predict that the
logarithmic 
singularity $\delta C_i\propto -\ln(|N-\frac{1}{4}|)$ should be
observed at any value of $t_1\neq 1$ (as nicely illustrated in Figs.~3 and 4). 

\subsection{$r_2\rightarrow 1$}

In the opposite limit $r_1\ll r_2\rightarrow 1$, the interdot
constriction considerably impedes the charge
spreading between the dots. $Q_2$ becomes an integer-valued operator 
describing electrons which tunnel into the dot 2
(Eq.(\ref{Coulomb2})) and
charge fluctuations in dot 1 closely resemble the ones of a 
{\bf single dot} which is
strongly-coupled to one lead:
\begin{equation}
H_b\propto r_1(-1)^{Q_2}\cos(\pi N)
\cos\left(\sqrt{2\pi}\phi_s(0)\right){\cal T}_{1x}. 
\end{equation}
The Kondo energy scale 
\begin{equation}
\Gamma_1=\frac{E_c\gamma}{\pi}(2r_1)^2\cos^2(\pi N),
\end{equation} 
is identical to the one of the single-dot problem\cite{Matveev}, 
and assuming $r_1\neq 0$, 
$Q_1$ becomes also {\it quantized}. 

The small term 
$t_2\Psi_{2\alpha}^{\dagger}(L)\Psi_{1\alpha}(L)S_2^+ +h.c.$ here mostly
produces slight charge fluctuations in the dot 2, and
$\delta C_2\propto -\ln(|N-1/2|)$. 

\subsection{$r_1\approx r_2\ll 1$}

For $r_1\approx r_2$, a strong opposition between the single-dot 
($Q_i$ is quantized for $r_2\gg r_1$)
and the composite-dot ground-state 
($\bar{Q}_i=Q/2$ for $r_1\gg r_2$) arises
giving a fascinating ``hybrid''
regime where the fractional plateaus become gradually destroyed by acquiring a
{\it positive} slope (Fig.~4); Close to $N=1/2$, exploiting 
Eqs.~(\ref{energies}) and (\ref{energy}),
we can approximate $(f({\cal R}_1)=\ln{\cal R}_1+{\it const.})$
\begin{equation}\label{comp}
\delta\bar{Q}_1\propto \left(N-1/2\right)\hbox{\Large{(}} 
{\cal R}_1 f({\cal R}_1)-{\cal
  R}_2\ln(|N-1/2|)\hbox{\Large{)}},
\end{equation}
then inducing an exotic ``3-peak'' capacitance profile; ${\cal
  R}_i=(r_i)^2$ (inset in Fig.~4). 
The central peak becomes more pronounced by slightly increasing
$r_2$, whereas the external peaks only depends on $r_1$ (as long as $r_2\ll
  1$). 

\subsection{$r_1\rightarrow 1$ and $r_2\ll 1$}

It is worthwhile to compare with the
case $r_1\rightarrow 1$ and $r_2\ll 1$. 
Here, $Q=\sqrt{2/\pi}\phi_c(0)=n$ must be an {\it integer-valued} 
operator which guarantees $\bar{Q}_1=n/2$. The fractional plateaus 
remain by decreasing the interdot coupling and only their widths 
progressively reduce: $
N^*(n=1)-N^*(n=0)=1/2-2\eta{\cal R}_2\ln(1/{\cal
  R}_2)$;
$\eta>0$ is a constant parameter. More precisely, for $N_1=N_2=N$, it
is easy to rewrite the backscattering term as:
\begin{equation}\label{pair}
\hskip -0.3cm
H_b=\frac{\sqrt{\gamma a E_c v_F}}{\pi a}4r_2\cos\hbox{\Large{(}}\frac{n\pi}{2}\hbox{\Large{)}}
\cos\left(\sqrt{2\pi}\phi_s(L)\right){\cal T}_{2x},
\end{equation}
to Eqs.~(\ref{Coulomb}),(\ref{tunnel}) which then produces a Kondo energy scale
$\Gamma_2= {E_c\gamma}(2r_2)^2\cos^2\hbox{\Large{(}}\frac{n\pi}{2}
\hbox{\Large{)}}/\pi$,
and then a relative energy shift $\delta E\propto 
{\cal R}_2\ln(1/{\cal R}_2)$ between
{\it even} and {\it odd} states\cite{Matv2,Halperin}. This engenders that the
positions of the capacitance (sub-)peaks (furnished by Eq.~1) are shifted as 
\begin{equation}\label{splits2}
N^*=(2n+1)/4+(-1)^n 
\eta{\cal R}_2\ln(1/{\cal R}_2). 
\end{equation}
The capacitance (sub-)peaks are 
not equally spaced anymore and 
progressively pair around the points $N_s^*=(2n+1)/2$ (Fig.~2).
Finally, we have checked that integer plateaus become more prominent:
$N^*(n=2)-N^*(n=1)=1/2+2\eta{\cal R}_2\ln(1/{\cal R}_2)$.

\section{Conclusion}
 
In closing, based on two-impurity two-channel Kondo models, 
we have presented a detailed
discussion on the evolution of the fractional plateaus as a function
of the hopping parameters $t_1$ and $t_2$ for a double-dot
coupled via a single-mode QPC to a reservoir. 

Again, for {\it perfect}
interdot transmission, Coulomb steps of size 1/2
occur for any finite backscattering between the lead and the left dot. When an 
electron enters the artificial molecule,
a charge $1$ is fluctuating back and forth between the 
two dots. 
We are hopeful that this can be observed via 
capacitance measurements\cite{note-time,note-times}.

Substantially decreasing
the interdot coupling inevitably restores the single-dot Coulomb blockade
and the capacitance peak period {\it doubles}. 

For strong coupling between the lead and
the left dot $(r_1\ll 1)$, we find a striking 
intermediate range $(r_2\approx r_1)$ 
where the fractional steps become progressively
unstable, i.e., show an increasing positive slope; this happens 
due to the strong competition between a {\it single-dot} and a
{\it composite-dot} ground state. On the contrary, when $r_1\rightarrow 1$, 
$Q$ must be quantized and $\bar{Q}_1=Q/2$; the fractional steps 
persist.

For {\it asymmetric} dots, e.g., with different gate-dot capacitances, we 
report that the Coulomb staircase with halved steps is gradually
altered.

{\bf Acknowledgments:}
This work was supported by the Swiss National Science Foundation and is 
currently supported by NSERC. The
author is grateful to R.~C. Ashoori and K.~A. Matveev for their comments.


\begin{thebibliography}{99}
\bibitem{note-times}
For $t_2\rightarrow 1$, the
typical time of those fluctuations is 
$t_c\sim\hbar/E_c$ (For $t_2\ll 1$, $t_c$ is much larger $\sim
(\hbar/E_c)e^{1/t_2}$).
\bibitem{note2}
For simplicity, charge is now normalized to $e$.
\bibitem{note33}
We may {\it equally} use $|\Phi>_i$ or
$(|\Phi>_i *\ |Q_i>)$, where
$|\Phi>_i$ is any state of the dot i with charge $Q_i$; the (charge) state
$|Q_i>$ must be viewed as an extra label to $|\Phi>_i$. When the tunneling 
process $t_2\Psi^{\dagger}_{2\alpha}(L)\Psi_{1\alpha}(L)$ takes place, e.g., 
$Q_2$ changes from 0 to 1; this can be emphasized by the
introduction of a pseudo-spin
operator: $|Q_2=1>\ =S_2^+|Q_2=0>$\cite{Matv1,karyn2}. 
This procedure gives another type of term (with 2 spin operators) 
$t_2 \Psi^{\dagger}_{2\alpha}(L)\Psi_{1\alpha}(L)S_2^+S_1^-$; but, this
is not ``renormalized'' at low energy and we ignore it.
Again, here the coupling $S_{1z}S_{2z}$ $(Q_1 Q_2)$ is 
negligible.
\bibitem{extra}
We could also introduce
two different wave functions for the two QPCs 
($\psi(0)\rightarrow \psi_1(0)$, $\psi(L)\rightarrow \psi_2(0)$ in Eq.(17)).
\end{thebibliography}
\end{document}